Body of paper goes here:

\documentstyle [preprint,aps]{revtex}

\begin{document}

\title{Conductance fluctuations
due to a single bistable scatterer
in a weakly connected conductor}

\author{Vladimir I.  Fal'ko $^{*,+}$ }

\address{ $^{*}$ Max-Planck-Institut f\"ur
Festk\"orperforschung,
Heisenbergstr.  1, 70569 Stuttgart, Germany \\\
$^{+}$ Institute of Solid State Physics,
Chernogolovka, 142432 Russia}

\date{}

\maketitle\

\begin{abstract}

We study the effect of a single bistable scatterer
on a conductance
of mesoscopic conductors in the crossover regime
between open and
closed systems.  After the resistance of the contact
between the
metallic grain and bulk electrode exceeds the
resistance of a grain,
the effect of a bistability is enhanced and takes
the universal form.
Its efficiency scales with the cross-section of the
bistable impurity
and of conductances of contacts but is independent
of this impurity
position inside a sample.  Mutual correlations
between
magneto-conductance fluctuations at two conductance
levels also look
universal.  Enhancement of the conductance
sensitivity to variations
of impurity configurations in samples with highly
resistive contacts
transforms the spectral density of a quantum excess
noise toward a
white-noise behavior and requires higher stability
of the sample in
order to observe mesoscopic effects in
direct-current measurements.

\end{abstract}


\newpage

\section{Introduction}

Mesoscopic devices of various modifications often
show pronounced
telegraph and other low-frequency noise
\cite{Telegraph}.  Some part
of this noise has been interpreted \cite{Noise} in
terms of the
presence of bi- or multi-stable configurations of
defects or of a slow
impurity diffusion.  The origin of this effect has a
quantum nature.
In the absence of any inelasticity and at low
temperatures, the
conductance of an individual sample depends on the
interference
pattern of diffusive electronic waves in it
\cite{UCF}, and a bistable
scatterer manifests itself by the change of phases
of coherent
multiply scattered electron waves which touch it
during the pass
through the sample.  As a result, the recharging of
even a single
impurity among an infinite number of them can
produce a finite
effect on the conductance \cite{Noise}.

Nevertheless, in devices with the form of a
micro-bridge (open
systems), this effect is rather limited.  Recharging
or displacement
of a single short-range impurity among a lot of them
(their density
is $n$) produces a conductance variation of the
order of $\langle\delta
g^2\rangle\sim (l^2n)^{-1}$
\cite{Noise} (we use quantum units $e^2/h$ to
measure all conductances
below and brackets stand for an averaging over
static impurity
configurations).  This cannot provide a complete
renewing of a
random potential realization from the point of view
of a resulting
transmission through the sample, since the number of
impurities $n
l^2$
per area of a squared mean free path $l$ is usually
great.  Further,
both the absolute value and auto-correlation
function of conductance
variations in a magnetic field are quite sensitive
to the bistable
defect position with respect to current contacts
\cite{Falko}.

The goal of the present work is to describe the
effect of a single
scatterer on properties of a nearly closed
mesoscopic conductor and,
in particular, to estimate, how much changes one
should produce in a
disordered cavity with low-conductance leads in
order to collect full
statistics of conductance fluctuations (CF) in it.
The latter
information can be useful both for the studies of
mesoscopic effects
in electronic microdevices \cite{Marcus,Marcus1} and
transmission
experiments with microwaves \cite{Analog} which
could be directed to
the check of recently predicted universal
distributions of transport
coefficients of chaotic microcavities
\cite{StatisticsSS}.  The
calculation below confirms an intuitive expectation
that, as the
mesoscopic conductor gets less and less connected to
bulk electrodes
(conductances $g_b$ of contacts decrease), the
influence of an individual
scatterer on the conductance increases.  The
crossover occurs when
the contact resistance, $g_b^{-1}$ exceeds that of a
metallic grain ($
g_c^{-1}$).
After this, CF's produced by a bistability take the
universal
asymptotic behavior specific to zero-dimensional
systems:  their
r.m.s.  value, $\langle\delta g^2\rangle\sim
g_b^{-1}(\sigma_i/\lambda^{
d-1}_F)$ becomes independent both of the
position of the bistable impurity and of the density
of other
scatterers, but depends on the impurity
cross-section $\sigma_i$ weighted by
the electron Fermi wavelength $\lambda_F$ and the
contact conductance.  The
normalized magneto-correlation function between
conductances related
to different impurity states also takes the
universal form resembling
that of magneto-fluctuations of the density of
states in a closed
chaotic billiard \cite{Simons}.  The above statement
is also partly
valid at relatively high temperatures.

\section{Conductance fluctuations in a weakly
connected diffusive conductor}

In order to study the mesoscopic conductance in the
crossover regime
between completely open \cite{UCF} and completely
closed
\cite{Efetov} chaotic systems, we consider the
circuit composed of a
diffusive coherent metallic conductor (with
conductance $g_c$) connected
to reservoirs by two equal, highly resistive
contacts (with individual
conductances $g_b$).  The complication of
calculations specific to the
Coulomb blockade regime can be avoided by suggesting
that the
conductance $g_b\gg 1$.  The same condition,
reformulated in terms of
energetic characteristics of levels, holds the
inequality $\Gamma\gg
\Delta\epsilon$
between the characteristic escape-associated width
of states in a box,
$\Gamma$, and their mean level spacing
$\Delta\epsilon$.  The latter relation is known to
be a condition to the applicability of the
perturbation theory analysis
\cite{Efetov}, so that in what follows we slightly
modify the
conventional diagrammatic calculations \cite{UCF}
into a form which
allows us to incorporate the effect of resistive
contacts.  We do this
in the fashion of earlier works
\cite{Bound,Multiconnected}.

In a piece of a disordered metal ($L\gg
l\gg\lambda_F$), the electron
density-density and phase-phase correlations obey
the diffusion law,
the same as the electron distribution function
$N(\epsilon )$, $(\partial_
t-D\vec{\partial}^2)N(\epsilon )=0$,
with the same diffusion coefficient $D$ and the same
kind of boundary
conditions.  The latter are the continuity and
current conservation
equations which one should read as $\vec
{n}D\vec{\partial }N=0$ at the wire boundaries
and $(\pm 2\alpha L)\partial_xN=N-N_F(\epsilon\pm
eV/2)$ at the edges $
x=\pm L/2$, respectively.
Here, $N_F$ is the Fermi distribution function, and
we already exploited
the quasi-1d wire geometry and separate the
longitudinal ($x$) and
transverse variables.  The meaning of the parameter
$\alpha$ can be
clarified after comparing the value of an averaged
current through
three resistors ($g_b^{-1}$, $g_c^{-1}$ and again
$g_b^{-1}$) in series with the value
calculated from the diffusion equations using the
definition of a total
current as an integral of a local current density in
a metallic part of
this circuit, $<I>=-eDL^{-1}\nu_F\int d{\bf
x}d\epsilon [\partial_
xN(\epsilon )]$ ($\nu_F$ is the density of states
at the Fermi level and $L$ is the length of a wire).
 After comparison,
we determine the parameter $\alpha$ as the ratio
between the wire to
contact conductances,

\begin{equation}
$$\alpha =g_c/2g_b,$$
\label{alfa}\end{equation}
so that an open system corresponds to
$\alpha\rightarrow 0$ and an quasi-isolated one
is described by the limit of $\alpha\gg 1$.

Using the above definition of the current through
the circuit in
terms of the integrated current density in its
metallic part, {\it the
correlation function $K=\langle g_1g_2\rangle
-\langle g\rangle^2$\/} of two conductances,
$g_{1(2)}=g(U\pm\delta U/2,H\pm\Delta H/2)$, taken
for different impurity
configurations $U_{1,2}=U\pm\delta U/2$ and at
different magnetic fields
$H_{1,2}=H\pm\Delta H/2$ can be related \cite{UCF}
to two-particle Green
functions $P_{d(c)}=\langle
G^R_{U_1H_1}G^A_{U_2H_2}\rangle$ averaged over
realizations of a
'background' random potential $U$, so called
diffusons and Cooperons.
The latter two describe the density-density and
phase-phase
correlations of electrons in a diffusive regime and,
after accounting
for the variation $\delta U$, they can be found from
the equations
\cite{Falko}

\begin{equation}
$$\{-i\Omega -D(\vec{\partial }-i{e\over {\hbar
c}}\vec {A}_{d(c)}
)^2+u\tau_f^{-1}V\delta (\vec {r}-\vec
{r}_i)\}P_{d(c)}(\Omega ,\vec {
r},\vec {r}^{\prime})=\delta (\vec {r}-\vec
{r}^{\prime}).$$
\label{Diff}\end{equation}
In this equation, the effect of a single bistable
scatterer (placed at
the coordinate $\vec {r}_i=(\xi_iL,{\bf
r}_{\perp})$) is incorporated into a $
\delta$-functional
repulsive potential with the strength
$$u\approx\left({L\over l}\right)^2{1\over
{Vn}}{{\sigma_i}\over {
\sigma_{tr}}},$$
where $Vn$ is the total number of short-range
scatterers inside the sample
with volume $L\times S_{\perp}$, $\sigma_{tr}$ -
their transport cross-section, $
\sigma_i$ - the
cross-section of a varied impurity potential $\delta
U$, and $\tau_
f=L^2/D$.  Eq.
(\ref{Diff}) should be completed with the boundary
conditions
$$\{\vec {n}(\vec{\partial }-i{e\over {\hbar c}}\vec
{A}_{d(c)})\}
P_{d(c)}=0$$
at the surface, and
$$P_{d(c)}=(\pm\alpha L)(\partial_x-i{e\over {\hbar
c}}A_{x,d(c)})
P_{d(c)}$$
at contacts to reservoirs, $x=\pm L/2$.  Here,
$\alpha$ has the same meaning
as above and one can see that the escape-induced
term in the
boundary conditions on the two-particle Green
function $P_{d(c)}$ is twice
more efficient than that of a single-particle
distribution function
$N(\epsilon ).$  For the sake of convenience, we
choose such a gauge of the
field $\vec {A}_{d(c)}$, $rot\vec {A}_{d(c)}$ $=\vec
{H}_1\pm\vec {
H}_2$, which provides us with $\vec {n}\vec {A}=0$
and
allows an easy transition to quasi-1D formulae.  In
particular, the
above equations can be applied to the lowest
transverse diffusive
mode and, therefore, we rewrite it in the form where
the presence
of a gauge field in a long derivative,
$(\vec{\partial }-i{e\over {
\hbar c}}\vec {A})^2$, is replaced by an
extra decay rate $\gamma^2$:  we get
$(\partial_x^2-\gamma^2/L^2)$, where
$\gamma_{d(c)}^2={{(2\pi )^2}\over
3}(\phi_1\pm\phi_2)^2$ is determined by the value of
a magnetic field
fluxes, $\phi_{1,2}=SH_{1,2}/\Phi_0$, through sample
area, and $\Phi_
0=hc/e$.

If we restrict the analysis to the case $g_b\gg 1$,
we can confine the
perturbation theory calculation of $K=\langle
g_1g_2\rangle -\langle
g\rangle^2$ to the diagrams
\cite{UCF} which contain two diffusons or Cooperons,
so that

\begin{equation}
$$K={{4D^2}\over {L^2}}\int d\epsilon
d\epsilon^{\prime}\int dxdx^{
\prime}[\partial_x\partial_{\epsilon}N(\epsilon
)][\partial_x\partial_{
\epsilon^{\prime}}N(\epsilon^{\prime})]\sum_{d(c)}\{
\mid P_{d(c)}(
\Omega ;x,x^{\prime})\mid^2+{1\over 2}\Re
P^2_{d(c)}(\Omega ;x,x^{
\prime})\},$$
\label{Def}\end{equation}
where $\Omega =\epsilon -\epsilon^{\prime}$. (The
diagrams of higher orders in $
P_{d(c)}$ are
suppressed by the factor $g_b^{-1}$).  After
substituting $N$ and $
P_{d(c)}$ from
Eq. (\ref{Diff}) into Eq.  (\ref{Def}), the latter
can be reduced to the
form $K(u,\phi_1,\phi_2)=$ $K_d+K_c=$
$K(u,\Delta\phi )+K(u,2\phi
)$,

\begin{equation}
$$K(u,\tilde{\phi })={6\over {(1+4\alpha
)^2}}\sum_{q_n}{1\over {[
q_n^2+\gamma^2(\tilde{\phi })]^2}}.$$
\label{K}\end{equation}

In deriving Eq.  (\ref{K}) and in finding the set of
$q_n$'s which stay in
this expression, we treat $P_{d(c)}$ as the Green
functions of equations
(\ref{Diff}) which are similar to the Schroedinger
equation of a
particle in a corresponding potential,
$$P(\Omega
;x,x^{\prime})=\sum_n\psi_n(x)\psi_n(x^{\prime})/(-i
\Omega
+\Omega_n).$$
Here, $\psi_n$ are the eigenfunctions and
$\Omega_n=Dq_n^2$ are the eigenvalues of
the spectral problem given by Eq.  (\ref{Diff}) with
its boundary
conditions.  For the case of a single bistable
scatterer in a quasi
one-dimensional wire placed at $x_i=\xi_iL$ we write
$\psi_n^{l(r)}=a_{l(r)}\cos
(q_nx+\theta_{l(r)})+b_{l(r)}\sin (q_n
x+\theta^{\prime}_{l(r)})$, where the indices $l(r)$
denote the wire intervals to the left (right) from
the scatterer $
\xi_iL$.
Using the analogy between the generalized diffusion
equation
(\ref{Diff}) and the Schroedinger equation, we
repeat the conventional
way \cite{Land} to find the spectrum of $q$'s.  That
is, the coefficient
$a$ and $b$, the phases $\theta^{\prime}$s and the
spectrum of $q$'s have to be found
from the above-mentioned boundary condition at the
wire edges and
after adjusting functions $\psi^l$ and $\psi^r$ at
the scatterer position:  $
\psi$ has
to be continuous in this point and the jump of its
derivative $\partial_
x\psi$ is
determined by the strength of the $\delta$-potential
in Eq.  (\ref{Diff}).
This gives one the set of linear equations.  After
eliminating the
coefficients $a$, $b$, $\theta$ and
$\theta^{\prime}$ from the set of linear equations,
we arrive
at the eigenvalue equation $F(q)=0$ on the number of
wave modes $q_
n$.

Skipping the details of this linear algebraic
procedure, we find that
for samples with the form of a single wire
$$F(q)=2\cos q+{{\sin q}\over {\alpha q}}-\alpha
q\sin q+u\alpha f
.$$
In this expression the function $f$ includes all the
necessary
parameters of a bistable scatterer,
$$f={{\sin q}\over {\alpha q}}+{{\cos q+\cos
((1-2\xi_i)q)}\over 2}
-{{\cos ((1-2\xi_i)q)-\cos q}\over {2(\alpha
q)^2}}.$$
In the case of a ring (Aharonov-Bohm) geometry
\cite{AhBosc}, the
function $F$ can be derived from the same set of
equation as before,
but with the boundary conditions on the electron
diffusion in two
semi-circlets (with a total conductance $g_c)$
connected near contacts.
It has the form
$$F(q)={{1-\cos (2q)}\over {(\alpha q)^2}}+\cos
(2q)-\cos (2\pi\tilde{
\phi })+{{2\sin (2q)}\over {\alpha q}}+{{\alpha
u}\over 2}f,$$
where
$$f={{\sin (2q)}\over {\alpha
q}}+\sum_{\pm}\left[{{\cos ((1\pm 2\xi_
i)q)-\cos (2q)}\over {(\alpha q)^2}}+{{\sin ((1\pm
2\xi_i)q)-\sin
(2q)}\over {(\alpha q)^3}}\right].$$
and $\tilde{\phi}$ denotes the magnetic field flux
encircled by a ring and
measured in units of $\Phi_0$.

The summation over the set of eigenvalues $q_n$ can
be performed
using the following procedure.  We rewrite the sum
in Eq.  (\ref{K})
as a result of an integration of the function
$R(z)=F^{\prime}(z)/(F(z)(z^2+\gamma^2)^2)$ over a
complex variable $
z$ along the
contour C 'encircling' the real axis.  Function
$R(z)$ has poles $
z=\pm q_n$
on the real axis, two poles $z=\pm i\gamma$ on the
imaginary axis and tends
to zero at $z\to\infty$.  Therefore, it is natural
to shift the contour C to
$\pm i\infty$, so that the value of the correlation
function $K$ would be
determined by the residues of $R(z)$ at $z=\pm
i\gamma$.  This gives us

\begin{equation}
$$K(u,\phi )={6\over {(1+4\alpha
)^2}}\Re\left\{{d\over {dz}}\left
({{-F^{\prime}(z)}\over {F(z)(z+i\gamma
)^2}}\right)_{z=i\gamma}\right
\}.$$
\label{General}\end{equation}

\section{Universal correlation function of
conductance fluctuations in
nearly closed systems}

Now one can describe the correlations
$K(u,\phi_1,\phi_2)$ at any value of
parameters involved and analyze the asymptotic
limits.  In particular,
we can follow the {\it features of
magnetoconductance fluctuations from
open\/} ($\alpha\to 0$) {\it to closed\/}
($\alpha\to\infty$) {\it systems\/}.  In an open
system, we arrive
at the well known r.m.s.  value of universal CF
\cite{UCF,StatisticsSS} and the correlation
function, $K(\phi_1,\phi_
2)=$
$K(2\phi )+K(\Delta\phi )$, with
$$K=\left\{\matrix{{3\over
2}{{2\gamma^2+2+\gamma\sinh (2\gamma )-\cosh
(2\gamma )}\over {\gamma^4[\cosh (2\gamma
)-1]}},\;wire\cr
{1\over {15}}\hbox{${{1+2\alpha^2{\rm c}{\rm o}{\rm
s}(2\pi\tilde{
\phi })}\over {[1-{1\over 2}\alpha^2{\rm c}{\rm
o}{\rm s}(2\pi\tilde{
\phi })]^2}}$},\;\;\;\;\;\;\;\;\;\;\;ring\cr}
\right.,$$
which describes both wire and ring geometries.

In an opposite limit of a weak connection
($\alpha\gg 1$), we get a different
geometrical prefactor in the r.m.s.  value of CF in
the unitary limit,
$var(g)=3/32$, and find the flux-dependent
correlation function of a
ring $K(\phi_1,\phi_2)=$ $K(2\phi )+K(\Delta\phi )$,

\begin{equation}
$$K(\tilde{\phi })={1\over {16}}\hbox{${{1+{1\over
2}{\rm c}{\rm o}
{\rm s}(2\pi\tilde{\phi })}\over
{\left[1+{{\alpha}\over 4}(1-{\rm c}
{\rm o}{\rm s}(2\pi\tilde\phi ))\right]^2}}$},$$
\end{equation}
which shows periodically repeated narrow splashes in
the form of
squared Lorentzians of a width $\Phi_0/\sqrt
{\alpha}$.

{\it The effect of a single bistable impurity on a
sample conductance \/}
can be analyzed from the dependence of the
correlation function $K$ on
the parameter $u$.  After studying the limit of
$\alpha\to 0$ (open system),
one can see that the effect of a single scatterer on
a conductance
can never be strong enough to renew completely the
realization of a
random potential configuration.  Second, the jump in
the conductance
value due to a change of one impurity shows a
dependence (through
the parameter $\rho ={1\over 4}-\xi_i^2$) on its
position relative to the wire edges,
both in the bistable scatterer efficiency,
$$1-{{K(u)}\over {var(g)}}=2\rho^2u{{2(1+\rho
u)+\rho (4-\rho u)}\over {
(1+\rho u)^2}},$$
and in auto-correlation properties under a variation
of a magnetic
field.  The latter could even allow one to make a
rough tomography
of a bistable impurity \cite{Falko}.

In a system with highly resistive contacts,
$\alpha\gg 1$, the electron spends
much longer time inside the mesoscopic conductor, as
compared to
the diffusion time $\tau_f=L^2/D$.  A classical
diffusive trajectory in a
'box' is much longer than in an open system, so that
it has better
possibility to meet a bistable scatterer many times
during different
traversals from one contact to another.  This
produces its higher
efficiency in renewing the sample realization and
results in the
universal form of the correlation function
$K(u,\phi_1,\phi_2)$,

\begin{equation}
$${{K(u,\phi_1,\phi_2)}\over
{var(g)}}=\left(1+{{\alpha}\over 2}[u
+\gamma_d^2(\Delta\phi
)]\right)^{-2}+\left(1+{{\alpha}\over 2}[u+
\gamma_c^2(2\phi )]\right)^{-2}.$$
\label{Correlations}\end{equation}
The first term in the right hand side describes
correlations between
$g_1$ and $g_2$ in the unitary limit.  The second
can be used for
discussing the crossover regime to the orthogonal
($H=0$) ensemble.
The parameter $u=\left({L\over l}\right)^2{1\over
{Vn}}{{\sigma_i}\over {
\sigma_{tr}}}$ related to a change of a scatterer
enters into this expression in combination with
enhancement
parameter $\alpha =g_c/2g_b$.  After using the
Drude-Einstein formula, we
find that the effect of a single scatterer is scaled
by the combination
of its scattering cross-section and the conductance
$g_b$,

\begin{equation}
$$\alpha u\approx{1\over {g_b}}{{\sigma_i}\over
{\lambda_F^{d-1}}}
.$$
\label{Parameter}\end{equation}
Remarkably, this combination doesn't depend on
properties of other
scatterers, except that we think about the limit of
$\lambda_F<l<L$.

Eqs.  (\ref{Correlations}, \ref{Parameter}) tell us,
how many
scatterers have to be changed/shifted in order to
collect the full
statistics of transmission fluctuations through a
quasi-closed quantum
box.  In electronic devices prepared of metals with
short-range
scatterers, the effect of single impurity is never
strong enough
until the system becomes completely closed and gets
into the Coulomb
blockade regime.  In low-density semiconductor
devices, the
displacement of a trapped charge near the 2D channel
affects electron
scattering phases more than a short-range defect can
do it.  A rough
estimation shows that, after 2D screening at the
distance of a
donor-related Bohr $a_B$ is taken into account, a
single recharging event
can be sufficient enough if $g_b\sim\lambda_F/a_B$.
As applied to microwave
experiments on a single mode transmission through
diffusive cavities
(that corresponds to $g_b=1$), Eqs.
(\ref{Correlations}, \ref{Parameter})
show that a single scatterer (such as a small piece
of a metal) can
already produce enough changes, if its size is of
about a wavelength
of irradiation.

The multiple diffusive traversal of a quasi-closed
system by an
electron - from contact to contact - destroys the
memory about a
specific impurity position, which makes the
magneto-tomography of a
bistable scatterer no longer possible.  This is
manifested by the fact
that the correlation function of random
magnetoconductance
variations takes the universal form Eq.
(\ref{Correlations}) similar
to that of thermodynamic parameters of isolated
systems
\cite{Simons}, but with a rescaled correlation
magnetic field flux
$\phi_c=H_cS$.  The field $H_c$ in a weakly
connected mesoscopic conductor
lies in between of what is known for open systems,
$\phi_c(\alpha\to
0)=\Phi_0${\it , \/}
and $\phi_c(\Gamma <\delta\epsilon )=\Phi_0\sqrt
{\delta\epsilon /
E_c}$ determined by the mean level spacing
$\delta\epsilon$ in
isolated metallic grains \cite{Efetov,Simons}.
Since the lifetime of a
diffusive electron inside a weakly connected
conductor (the quantity
inverse to $\Gamma /h$) is $\alpha\gg 1$ times
longer than the diffusive flight time
$\tau_f=L^2/D$, the mean square
$\langle\phi^2\rangle$ of a magnetic field flux
encircled by
a characteristic diffusive chaotic trajectory is
$\alpha$ times greater than
the geometrical flux $SH$ through the sample area.
This rescales
\cite{Multiconnected,Billiard,Efetov1} the
correlation magnetic field
flux of CF, in our case - down to the value of
$\phi_c=\Phi_0/(\pi\sqrt {
\alpha})\ll\Phi_0$.

In a weakly connected conductor, {\it a finite
temperature\/} shows up at
the scale of $2\pi T\ge\Gamma
=h\tau^{-1}_f/\alpha\sim E_c/\alpha$.  Thermal
smearing of the
Fermi distribution function beyond a shortened
correlation energy
$\Gamma =\alpha^{-1}h\tau_f^{-1}$ partially cancels
CF coming from 'independent' spectral
intervals.  At those high temperatures, the
correlation function $
K$ can
be calculated from Eq.  (\ref{Def}) by transforming
the integral over
energies into the sum over Matsubara's frequencies,
and at $\alpha
\gg 1$ we
arrive at

\begin{equation}
$${{K(u,\Delta\phi )}\over {var(g)}}={{\Gamma}\over
{9T}}\left(1+{{
\alpha}\over 2}[u+\gamma_d^2(\Delta\phi
)]\right)^{-1}$$
\label{KT}
\end{equation}
which again has universal parametric dependence
\cite{RemFourier}.

Since in earlier experiments
\cite{Marcus,Marcus1,Keller} the
magneto-correlations of conductance fluctuations
were also studied
using the Fourier transform representation of the
datas, it is
reasonable to compare the mean square values of the
Fourier
components of conductance fluctuations in a magnetic
field with what
one can get from Eqs.  (\ref{Correlations}) and
(\ref{KT}).  R.m.s.
values of a Fourier transform of AB oscillations in
a ring with
respect to the flux $\phi$ says directly about
efficiency of the electron
escape from a sample:  $<\delta
g^2_k>\propto\left(\alpha /2\right
)^{2k}$, $\alpha\ll 1$.  In a weakly
connected sample, $\alpha\gg 1$,
$$<\delta g^2_k>={{var(g)}\over {\sqrt
{2\alpha}}}\left(1-{{2\sqrt
2}\over {\sqrt {\alpha}}}\right)^k\kappa (T/\Gamma
),$$
where the most serious temperature dependence of the
'spectral
function' is incorporated into the multiplyer
$\kappa$.  In two limits of a
low and high temperatures, $\kappa (0)=1$ and
$\kappa (T\gg\Gamma
)={{2h\tau^{-1}_f}\over {9T\alpha}}$.
\cite{RemFourier} This behavior is consistent with
the r.m.s value of
the Fourier transform of CF with respect to the flux
$2\pi\phi$ through
the sample of a single-connected geometry:  At
$\alpha\gg 1$ we get

\begin{equation}
$$<|\delta g(k)|^2>=var(g){{\pi}\over b}{{\kappa
(T/\Gamma )}\over {\sqrt {
2\alpha}}}e^{-bk\sqrt 2/\sqrt {\alpha}},$$
\label{Fourier}
\end{equation}
where $b\sim 1$ is a geometrical factor.  The factor
$\kappa (T/\Gamma
)$ is the same
as above. One can see that the exponential
dependence of the
Fourier transform on the 'frequency' $k$ of random
magneto-conductance
oscillations is similar for the low and high
temperatures, like it has
been observed in \cite{Marcus,Marcus1}.

\section{Bistable-scatterer-induced excess noise in
a quantum conductor}

Higher sensitivity of nearly closed systems to a
change of the
impurity potential causes modifications of the
spectral shape of the
{\it low-frequency excess noise\/} (generated by
random switches of
several, $\delta N_i\gg 1$, localized charges or
defects) towards the
white-noise dispersion.  That is, a variation of
scatterers at the
characteristic time-scale $\tau_r$, $\langle
U(0)U(t)\rangle\propto
\kappa (t/\tau_r)$, related to a spectral
dispersion of the source of the noise at the
frequency $\omega_r\sim
\tau_r^{-1}$
destroys the correlation of instant conductance
values,
$K(t)\sim\left(1+{{\alpha u}\over 2}(1-\kappa
(t/\tau_r))\right)^{
-2}$, at a much shorter time scale:
$t_{*}=\tau_r\tau_i/(\alpha\tau_f)$.  This
estimation can be produced in the same way as
in Ref.  \cite{Spin}.  In order to do this, we
introduce the scattering
rate $\tau_i^{-1}=u\tau_f^{-1}\propto\delta N_i$
instead of the parameter $
u$.  The rate $\tau_i^{-1}$ says
us how often the diffusive electron scatters on a
bistable impurity,
so that the ratio $\alpha\tau_f/\tau_i\gg 1$ gives
the probability that an electron meets
this scatterer during his life-time inside a sample.
 Therefore, the
time $t_{*}$ can be obtained from the condition
$(t_{*}/\tau_r)(\alpha
\tau_f/\tau_i)\sim 1$ which
requires that each diffusive path can meet a renewed
scatterer with
a unit probability, and this will be the value which
determines the
spectral shape of a low-frequency excess noise in a
current.  By the
definition, the latter is $S(\omega )=\int
dte^{i\omega t}\langle
I(t)I(0)\rangle$ and can be found as

\begin{equation}
$$S(\omega )\sim\left({{e^2V}\over
h}\right)^2{{\tau_r\tau_i}\over {
\alpha\tau_f}}\left\{\matrix{1,\;\;\:\omega
<1/t_{*}\cr
\left(\omega t_{*}\right)^{-2},\;\:\omega
>1/t_{*}\cr}
.\right.$$
\label{S}\end{equation}
{}From this, as more breathing impurities are located
inside a metallic
region, the spectral density of a noise is
redistributed over a
range of frequencies $\sim 1/t_{*}$ much wider than
spectral range of the
source of a noise (the noise is transformed to the
white noise). It's
also clear that the integral intensity $\int d\omega
S(\omega )=\delta
I(t)^2\sim (e^2V/h)^2$
remains fixed.

On the other hand, the above consideration shows
that the
observation of effects of coherence in
direct-current measurements
in a closed system requires a higher stability of
samples, as
compared to open ones.  Any internal source of a
soft noise
reduces the amplitude of mesoscopic CF down to the
value
$var(g)=K(t\to\infty )$.  This would be, for
example, crucial for their
observation in (even slightly) magnetically
contaminated systems.
That is, the electron {\it spin-flip scattering\/}
makes conductance
dependent on an instant configuration of spins of
few magnetic
impurities, whereas the Korringa relaxation of
impurity spins -
caused by the same flip-flop with thermal or current
electrons -
varies the transmission in time \cite{Spin,Spin1}.
This effect is
completely analogous to the impurity recharging or
displacement
discussed above and leads to the same result:  The
mesoscopic
dc-conductance fluctuations are easily washed out
even when the
spin-flip scattering length in a bulk material,
$l_i=(D\tau_i)^{1/
2}$, is much
longer than sample dimensions.  Therefore, at low
magnetic fields,
dc-fluctuations are suppressed down to the r.m.s.
value

\begin{equation}
$$var(g)\sim\left({{\tau_i}\over
{\alpha\tau_f}}\right)^2=\left({{
l_i}\over {L\sqrt {\alpha}}}\right)^4\ll 1.$$
\end{equation}
A spectacular feature of a paramagnetic impurity
system can be
expected after application of a high enough magnetic
field.  If the
latter is able to polarize spins of scatterers,
their time-dependent
variation would be suppressed which can be observed
as an abrupt
restoration of the universal CF's in
dc-measurements.

\section{Conclusions}

In a summary, we have analyzed the conductance
fluctuations in a
mesoscopic piece of a metal in the crossover regime
from an open to
a nearly closed system.  On the base of the
perturbation theory
calculations we conclude that after the resistances
$g_b^{-1}$ of the device
contacts to bulk electrodes exceed the resistance
$g_c^{-1}$ of a metallic
grain itself, the sensitivity of a sample
conductance to the variation
of a bistable scatterer increases, so that at
$g_b\sim 1$ even a single
scatter can produce a variation $\delta g\sim 1$.
In the same limit, the
correlations function of conductance fluctuations
takes the universal
form specific to the zero-dimensional system, both
with respect to
the variation of an impurity configuration and
external parameters,
such as a magnetic field.

Although the calculations are performed in the
diffusion regime, the
derived above magneto-correlation functions seem to
be able to
describe fluctuations in chaotic ballistic systems.
In particular, Eqs.
(\ref{Correlations},\ref{KT},\ref{Fourier}) are in a
good agreement
with observations of Refs.
\cite{Marcus,Marcus1,Keller}.  One can
also expect that in billiards the variation of an
impurity can be
replaced by the variation of a shape, but before
using the derived
above equations one has to take into account the
following remark.
The variation of a shape is usually produced by some
variation of a
voltage $\delta V_g$ applied to side gates
\cite{Marcus}.  This variation of
gate voltages changes not only the shape of a
structure, but also its
area, $S\to S+\delta S$.  In an arbitrary case, the
change of an area occurs
in the first order on $\delta V_g$, $\delta
S\propto\delta V_g$, and results in a proportional
rescaling of the energies $\epsilon_m$ of all
single-particle states in the box
spectrum:  $\epsilon_m\to\epsilon_m/(1+\delta S)$.
Therefore, the states near the Fermi
level flow together through the Fermi level roughly
preserving their
spatial structure, and, therefore, their transport
abilities, since the
change of a scattering due to a direct shape
variation can appear
only as a quadratic term on $\delta V_g$.  This
results in the same effect as
if one would vary the Fermi energy itself
\cite{Keller} and has to be
described by the correlation function
$${K\over {vag(g)}}=\left(\left[1+{{\alpha}\over
2}\gamma^2(\Delta
H)\right]^2+\left[{{\alpha\tau_f\delta\epsilon}\over
h}\right]^2\right
)^{-1},$$
where $\delta\epsilon\propto\delta S\propto\delta
V_g$.  It's amusing to mention that, after necessary
substitutions, the parameter
$\alpha\tau_f\delta\epsilon /h$ which stands in the
function $
K$
can be rewritten as $\delta N_e/g_b$, where $\delta
N_e$ is an actual change of a
number of carriers assigned to the interior of a
metallic box.

This work has been stimulated by discussions with
J.C.  Maan of
experiments on the transmission through disordered
microwave
cavities.  I also thank K.B.  Efetov for comments on
universality in
zero-dimensional systems and an information
concerning his work
\cite{Efetov1} and C.  Marcus for discussions of the
shape-variation
effects in ballistic quantum billiards
\cite{Marcus1}.  A partial
support from ISF Grant REE-000 and NATO CRG 921333
is acknowledged.


\begin{references}

\bibitem{Telegraph}
M.J. Kirton and M.J. Uren, Adv. Phys. {\bf 38}, 367
(1989);
M.B. Weismann, Rev. Mod. Phys. {\bf 60}, 537 (1988);
{\it ibid.\/} {\bf 65}, 829 (1993) and
refs. therein.

\bibitem{Noise}
B.L. Altshuler and B.Z. Spivak, JETP Lett.  {\bf
42}, 447 (1985);
S. Feng et al, Phys.  Rev. Lett. {\bf 56}, 1960
(1986)

\bibitem{UCF}
B.L. Altshuler, JETP Lett.  {\bf 41}, 648 (1985);
P.A. Lee and A.D. Stone, Phys.  Rev.  Lett.  {\bf
55}, 1622 (1985);
A.D. Stone, Phys.  Rev.  Lett.  {\bf 54}, 2692
(1985);
B.L. Altshuler and D.E. Khmel'nitskii, JETP Lett.
{\bf 42}, 359 (1985);
D.E. Khmel'nitskii and A.I. Larkin, Physica Scripta
T {\bf 14}, 4 (1986)

\bibitem{Falko}
V.I. Fal'ko and D.E. Khmel'nitskii, JETP Lett. {\bf
51}, 189 (1990);
G. Bergmann, Phys. Rev. B {\bf 49}, 8377 (1994)

\bibitem{Marcus}
C.M. Marcus et al, Phys. Rev. Lett. {\bf 69}, 506
(1992);
C.M. Marcus et al, Phys. Rev. B {\bf 48}, 2460
(1993)

\bibitem{Marcus1}
I.H. Chan, R.M. Clarke, C.M. Marcus, K. Campman and
A.G. Gossard,
preprint;
R.M. Clarke, I.H. Chan, C.I. Duruoz, J.S. Harris,
C.M. Marcus, K. Campman and A.G. Gossard,
preprint

\bibitem{Analog}
H.J. Stockmann and J. Stein, Phys. Rev. Lett. {\bf
64}, 2215 (1990);
E. Doron et al, {\it ibid.\/} {\bf 65}, 3072 (1990);
H.-D. Graf et al, {\it ibid.\/} {\bf 69}, 1296
(1992);
A. Kudrolli et al, Phys. Rev. E {\bf 49}, 11 (1994)

\bibitem{StatisticsSS}
V.N. Prigodin et al, Phys. Rev. Lett. {\bf 71}, 1230
(1993);
C.W.J. Beenakker and B. Rejaei, Phys. Rev. Lett.
{\bf 71}, 3689 (1993);
J.T. Chalker and A.M.S. Macedo, ibid. 3693 (1993);
H.U. Baranger and P.A. Mello,  Phys. Rev. Lett. {\bf
73}, 142 (1994)

\bibitem{Efetov}
K.B. Efetov, Adv. Phys. {\bf 32}, 53 (1983)

\bibitem{Simons}
A. Szafer and B.L Altshuler, Phys. Rev. Lett. {\bf
70}, 587 (1993);
B.D. Simons and B.L Altshuler, Phys. Rev. Lett. {\bf
70}, 4063 (1993); Phys.
Rev B {\bf 48}, 5422 (1993)

\bibitem{Bound}
B.Z. Spivak and D.E. Khmel'nitskii, JETP Lett. {\bf
35}, 412 (1982);
P. Santhanam, Phys. Rev. B {\bf 39}, 2541 (1989)

\bibitem{Multiconnected}
R.A. Serota et al, Phys.  Rev. B {\bf 35}, 5031
(1987);
V.I. Fal'ko, Journ Phys. CM {\bf 4}, 3943 (1992)

\bibitem{Land}
L.D. Landau and E.M. Lifshitz, {\it Quantum
Mechanics\/}, Pergamon Press,
Oxford 1965, p. 65

\bibitem{AhBosc}
Y.  Gefen et al, Phys. Rev. Lett.  {\bf 52}, 129
(1984);
M.  Buttiker et al, Phys. Rev. A {\bf 30}, 1982
(1984)

\bibitem{Billiard}
R.A. Jalabert et al, Phys. Rev. Lett. {\bf 68}, 3468
(1992)

\bibitem{Efetov1}
K.B. Efetov, submitted to Phys. Rev. Lett.

\bibitem{RemFourier}
At high temperatures and $\alpha\gg 1$,

$K_{ring}(\phi )=$ ${{\pi\Gamma}\over
{96T}}\left(1+{{\alpha}\over
4}(1-{\rm c}{\rm o}{\rm s}(2\pi\phi ))\right)^{-1}$.

\bibitem{Keller}
M.W. Keller et al, Surface Science {\bf 305}, 501
(1994)

\bibitem{Spin}
A.A. Bobkov, V.I. Fal'ko and D.E. Khmel'nitskii,
JETP {\bf 71}, 393 (1990)

\bibitem{Spin1}
V. Chandrasekhar et al, Phys.  Rev. B {\bf 42}, 6823
(1990)

\end{references}
\end{document}